\documentclass{aa}
\usepackage{graphics}
\usepackage{times}

\newcommand{\lsi}    {LS~I+61$^{\circ}$303}
\newcommand{\ltsima} {$\; \buildrel < \over \sim \;$}
\newcommand{\simlt}  {\lower.5ex\hbox{\ltsima}}            
\newcommand{\gtsima} {$\; \buildrel > \over \sim \;$}
\newcommand{\simgt}  {\lower.5ex\hbox{\gtsima}}            

\begin{document}

\thesaurus{06     
        (08.09.2  
         08.05.2  
         13.18.5  
         13.25.5  
         ) }

  \title{Evidence of H$\alpha$ periodicities in \lsi}

  \author{R.K.~Zamanov\inst{1}
  \and J.~Mart\'{\i}\inst{2}
  \and J.M.~Paredes\inst{3}
  \and J.~Fabregat\inst{4}
  \and M.~Rib\'o\inst{3}
  \and A.E.~Tarasov\inst{5}}

  \offprints{R.K.~Zamanov, rozhen@mbox.digsys.bg}

  \institute{National Astronomical Observatory Rozhen,
              P.O.Box 136, BG--4700 Smoljan, Bulgaria
  \and Departamento de F\'{\i}sica, Escuela Polit\'ecnica Superior,
       Universidad de Ja\'en, C/ Virgen de la Cabeza, 2, E--23071 Ja\'en, Spain
  \and Departament d'Astronomia i Meteorologia, Universitat de
       Barcelona, Av. Diagonal 647, E--08028 Barcelona, Spain
  \and Departamento de Astronom\'{\i}a, Universidad de Valencia,
       E--46100 Burjassot, Valencia, Spain
  \and Crimean Astrophysical Observatory, 334413 Nauchny, Crimea, Ukraine}

\date{Received / Accepted}

\maketitle


\begin{abstract}

We present the results of analyzing H$\alpha$ spectra of the radio
emitting X-ray binary \lsi. For the first time, the same 26.5 d radio
period is clearly detected in the H$\alpha$ emission line. Moreover, the
equivalent width and the peak separation of the H$\alpha$ emission line
seem also to vary over a time scale of 1600 days. This points towards the
$\sim4$ yr modulation, detected in the radio outburst amplitude, being
probably a result of variations in the mass loss rate of the Be star
and/or density variability in the circumstellar disk. In addition, the
dependence of the peak separation from the equivalent width informs us
that the \lsi\ circumstellar disk is among the densest of Be-stars.


\end{abstract}

\section{Introduction}

The Be/X-ray binary \lsi\ (\object{V615 Cas}) is the optical counterpart
of the highly variable radio source \object{GT~0236+610}, which exhibits
radio outbursts every $26.4917\pm 0.0025$ d assumed to be the orbital
cycle (Taylor \& Gregory 1984; Gregory et al. 1999). The radial velocity
measurements of Hutchings \& Crampton (1981) are consistent with this
interpretation and suggest an eccentric orbit ($e \simeq 0.6$). The same
26.5 d period has been detected in the UBVRI photometric observations of
Mendelson \& Mazeh (1994), in the infrared (Paredes et al. 1994) and in
the soft X-ray flux (Paredes et al. 1997). The amplitude of the periodic
variations is \simlt0.2 magnitudes in the optical and infrared, while the
X-ray luminosity of the system often evolves within the range $10^{33}$ to
$6 \times 10^{34}$ erg s$^{-1}$ (Bignami et al. 1981; Paredes et al.
1997). \lsi\ is also currently regarded today as a serious counterpart
candidate to the $\gamma$-ray source \object{2CG~135+01} (Gregory \&
Taylor 1978; Kniffen et al. 1997; Strickman et al. 1998).

The \lsi\ radio emission has been interpreted as synchrotron radiation
from relativistic particles. The dependence of the radio outburst flux
density on frequency, the time delay in the outburst peak, and the general
shape of the radio light curves can be modeled as continuous injection of
relativistic particles into an adiabatically expanding plasma cloud or
plasmon (Paredes et al. 1991). It has been proposed that the genesis of
such plasmons could be the result of the transition of the neutron star
companion from the Propeller to the Ejector regime (Zamanov 1995).
Alternatively, a young radio pulsar may also be responsible for the radio
emission (Maraschi \& Treves 1981; Tavani 1994). The ejector regime would
then apply over the whole orbital period, with the flaring radio emission
resulting from the variable position of the shock front between the Be
star wind and the relativistic wind of the pulsar. Supercritical accretion
models during the periastron passage (e.g. Taylor et al. 1992) were also
widely accepted in the recent past. However, they appear now very
difficult to reconcile with the confirmed sub-Eddington X-ray luminosity
of \lsi.

On the other hand, a strong modulation in the amplitude of the 26.5 d
periodic radio outbursts, on a time scale of $\sim4$ yr, was noticed
independently by Paredes (1987) and Gregory et al. (1989). This modulation
has been recently confirmed after a Bayesian analysis of 20 yr of \lsi\
radio data by Gregory (1999). Two possible scenarios have been considered
in order to account for such a long-term variability: the precession of
relativistic jets or, alternatively, the variable accretion rate due to
quasi-cyclic Be-star envelope variations (Gregory et al. 1989).

The optical spectrum of \lsi\ corresponds to a rapidly rotating B0V star
with a stable equatorial disk and mass loss (Hutchings \& Crampton 1981).
H$\alpha$ observations of this star have been discussed, among others, by
Gregory et al. (1979), Paredes et al. (1994) and Zamanov et al. (1996). In
these papers, it was noticed a significant variability of the red and blue
humps at opposite locations of the orbit. On the other hand, the quasi
simultaneous H$\alpha$ and radio observations by Zamanov et al. (1996) did
not indicate considerable changes in H$\alpha$ close to the radio maximum.
This fact is in agreement with the assumption that the H$\alpha$ emission
comes mainly from the Be star disk, and the radio outbursts are related to
the compact object. Another interesting property of the H$\alpha$ emission
is its very broad wings, achieving $FWZI \sim 3000$ km s$^{-1}$.
Unfortunately, most \lsi\ spectra available do not have sufficient SNR to
investigate this intriguing feature.

In this paper, we present 84 new H$\alpha$ observations of \lsi\ which are
analyzed together with the data from Paredes et al. (1994) and Zamanov et
al. (1996). We report the clear detection of a 26.5 d periodicity in some
parameters of the H$\alpha$ emission line. In addition, our extended
spectroscopic database interestingly suggests the presence of a 1600 d
modulation in the H$\alpha$ equivalent width. These new H$\alpha$ periods
seem to match closely those known at radio wavelengths. The possible
implications of all these findings are pointed out and discussed.

\section{Observations}

Our H$\alpha$ spectroscopic observations of \lsi\ were carried out using
different astronomical facilities. The results of this observations are
summarized in Table 1.

A total of 60 new spectra were obtained with the Coud\'e spectrograph of
the 2.0 m RCC telescope at the Bulgarian National Astronomical Observatory
``Rozhen", between October 1993 and August 1998. An ISTA-CCD with
$400\times580$ pixel was used as a detector. The spectrum coverage is
about $110$ \AA, with a dispersion of $0.2$ \AA\ pixel$^{-1}$. Usually 2
or 3 exposures, of 20 minutes each, were always taken to improve the
sensitivity and to avoid problems with eventual cosmic ray hits. The
``Rozhen'' observations have been processed using the 3A software (Ilyin
1997) and the IRAF package of NOAO (Tucson, Arizona). It deserves noting
that the new Rozhen data from Ro931031 to Ro980805 are a uniform data set.

We have also used 17 more spectra obtained within the framework of the
Southampton and Valencia Universities collaborative program on high-mass
X-ray binaries (Coe et al. 1993; Reig et al. 1997a). 14 of them were
obtained with the Richardson-Brearly Spectrograph at the Cassegrain focus
of the Jakobus Kaptein Telescope, La Palma, Spain. Different grattings
were used, giving dispersions between 0.33 and 1.34 \AA\ pixel$^{-1}$. The
remaining 3 spectra were obtained with the P-60 Cassegrain Echelle
Spectrograph in regular gratting mode at 0.8 \AA\ pixel$^{-1}$, attached
to the 1.5-m telescope of the Palomar Mountain Observatory. These data
were reduced using the $Figaro$ (Shortridge \& Meyerdicks 1996) and
$Dipso$ (Howarth et al. 1996) packages included in the STARLINK software
collection.

Additionally, we also include 7 H$\alpha$ spectra from the 2.6 m telescope
at the Crimean Astrophysical Observatory. They were obtained at the
Coud\'e focus of this instrument with a dispersion of $0.066$ \AA\
pixel$^{-1}$ and a coverage of about $67$ \AA. The detector used was a
Photometrics SDS-9000 with a EVV 15-11 $1024\times256$ pixel CCD.

In order to ensure a unified post-processing of the spectra, all of them
were normalized with respect to the local continuum. When possible the
same continuum regions were used at 6520-6530 and 6595-6605 \AA, located
at the blue and red side of the H$\alpha$ emission line, respectively. The
underlying continuum was rectified to unity by employing a linear fit.

On each spectrum, we have measured the total equivalent width of the
H$\alpha$ emission line (hereafter $EW($H$\alpha)$), the heliocentric
radial velocities of the central dip, blue and red humps, the ratio
between the equivalent widths of the blue and red humps, the $B/R$ ratio
and the $FWHM$ of the humps. The dates of observations and the measured
quantities are summarized in Table 1. The $FWHM$ and the radial velocities
were measured by employing a Gaussian fit. The $FWHM$ was corrected for
the instrumental broadening. The errors depend mostly on the dispersion,
and we estimate them to be about 10\% for the $EW$s, about 1/2 pixel for
the radial velocities, $\pm0.04$ in the $B/R$ ratios and 1 pixel for the
$FWHM$s.

Some previously published Rozhen spectra, those from Ro920903 to Ro 930909
in Table 1, have been processed again to be used in this work. We will
also include in our analysis the H$\alpha$ parameters listed in the Table
4 of Paredes et al. (1994), whose $EW(R)/EW(B)$ header should be actually
the inverse.

\section{Results}

\subsection {A very dense circumstellar disk in \lsi}

It is well known that the H$\alpha$ peak separation and the equivalent 
width correlate for the Be stars. Hanuschik et al. (1988) have derived the 
law
  \begin{equation}
      \log \left[ \frac{\Delta V}
                {2\,v\,\sin{i}} \;\right]
       =\; -a\; \log \left[ \frac {EW({\rm H}\alpha)}{\rm \AA}\;\right] +\;b,
  \label{EW-peaks}
  \end{equation}
where $\Delta V$ is the peak separation, $v \sin{i}$ is the projected
rotational velocity and $EW($H$\alpha)$ is expressed in \AA. For stars
with $EW($H$\alpha)>3$ \AA, the average values are $a \simeq 0.4$ and $b
\simeq -0.1$. In the limit case of the most dense envelopes $b=+0.2$.
Since for rotationally dominated profiles $\Delta V/(2\,v\,\sin{i})\;$ can
be regarded as a measure of the outer radius inverse of the H$\alpha$
emitting disk (Hanuschik 1988), the law in Eq.~\ref{EW-peaks} simply
expresses that the outer radius grows as the $EW($H$\alpha)$ becomes
larger.

\begin{figure}[htb]
\mbox{}
\vspace{5.2cm}
\includegraphics{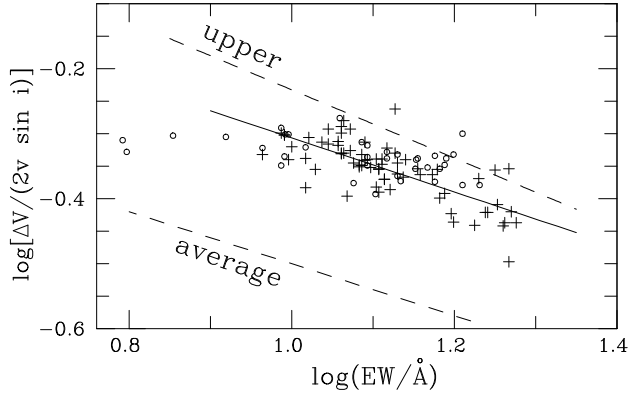}
\caption[]
     {Plot of $\log{(\Delta V/(2v\:\sin{i}))}$ versus $EW($H$\alpha)$ of
\lsi. The crosses represent the Rozhen data, the circles the other. The
solid line is the best linear fit over the Rozhen data. The dashed lines
are the average behavior and the upper limit for the Be stars. It is
visible that all data points are shifted towards denser Be disks}
\label{FdV}
\end{figure}

The same quantities for \lsi\ are plotted in Fig.~\ref{FdV} adopting
$\;v\,\sin{i}=360$ km s$^{-1}$ (Hutchings \& Crampton 1981). The solid
line represents the best linear fit over Rozhen data, from which we
derive: $a=0.38 \pm 0.04$ and $b=0.07 \pm 0.04$. The dashed lines
represent the average law $(a=0.4,\; b=-0.1)$ and the upper limit for the
Be stars as obtained by Hanuschik et al. (1988). Using their formulation,
our results imply that the circumstellar disk in \lsi\ is among the
densest disks in the Be stars, i.e., roughly a factor of 2 denser than the
average.

Dense circumstellar disks, optically thick at infrared wavelengths, have
been recently identified in other Be/X-ray binaries (\object{V0332+53},
Negueruela et al. 1999; \object{4U 0115+63}, Negueruela et al., in
preparation). These works suggest that the disks around Be stars in X-ray
binaries are denser and smaller that those in isolated Be stars. The
results of Reig et al. (1997b) point towards neutron star preventing the
formation of an extended disk in systems with short orbital periods,
presumably due to tidal truncation (Okazaki 1998).

\begin{figure}[htb]
\mbox{}
\vspace{6.9cm}
\includegraphics{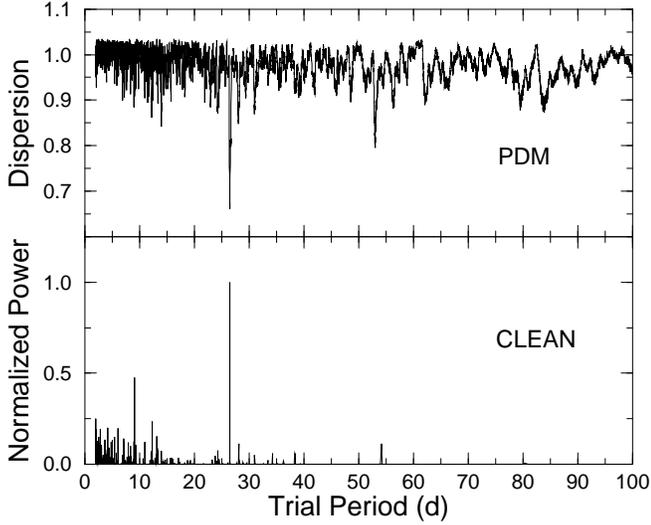}
\caption[]
     {Periodograms for the ratio $EW(B)/EW(R)$ in the H$\alpha$ emission
line of \lsi\ computed using the PDM (top) and CLEAN (bottom) algorithms.
The 26.5 d period is independently detected in both cases as the most
significant one}
\label{PDM26}
\end{figure}

\begin{figure}[htb]
\mbox{}
\vspace{16.0cm}
\includegraphics{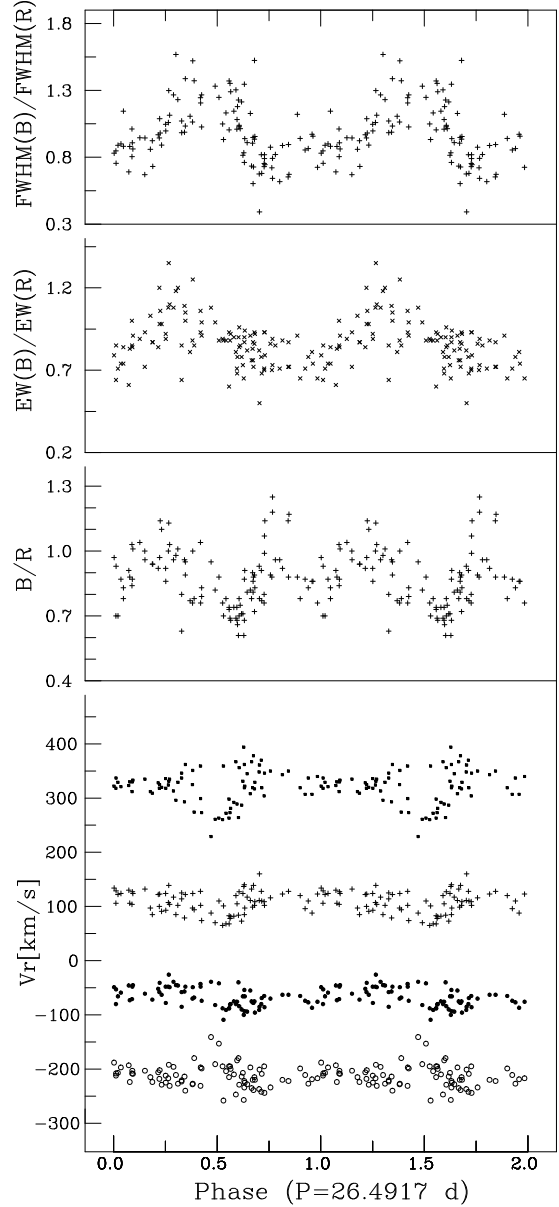}
\caption[]
     {H$\alpha$ line parameters of \lsi\ folded on the 26.4917 day period
and zero phase set at JD2443366.775. The panel of radial velocities
contains, from the lower to the upper plot, the heliocentric radial
velocities of the blue hump, $V_r(B)$, the central dip, $V_r(dip)$, the
red hump, $V_r(R)$, and the hump separation, $\Delta V$, respectively}
\label{P26}
\end{figure}


\subsection{Detection of 26.5 day period in H$\alpha$}

We have conducted a period analysis for the different H$\alpha$ line
parameters listed in Table 1. The period search methods used were both the
Phase Dispersion Minimization (PDM) (Stellingwerf 1978) and the CLEAN
algorithm (Roberts et al. 1987). As a result, we find clear evidences that
the H$\alpha$ emission in \lsi\ also displays variations with the same
26.5 d radio period. This behavior is better revealed in the line ratios
such as $FWHM(B)/FWHM(R)$, $EW(B)/EW(R)$ and $B/R$. The periods detected
are about $26.50-26.55$~d with typical error about $\pm0.04$ d. In
Fig.~\ref{PDM26} we present the PDM and CLEAN periodograms for the
$EW(B)/EW(R)$ case. In this representative example, the most significant
period detected in the range $2-100$~d corresponds to $26.51\pm0.03$ d and
$26.53\pm0.04$ d for the PDM and CLEAN methods, respectively. The first
sub-harmonic at 53 d is also clearly noticeable in the PDM panel. The
periodograms for the remaining line parameters, such as radial velocities
and equivalent widths alone, may also contain some power at the 26.5 d
period. However, in most cases the detections are still too noisy and
additional observations will be needed to confirm them.

In any case, Fig.~\ref{PDM26} can be considered as a reliable proof that
the signature of the \lsi\ radio period is present in the H$\alpha$ line.

\begin{figure}[htb]
\mbox{}
\vspace{10.6cm}
\includegraphics{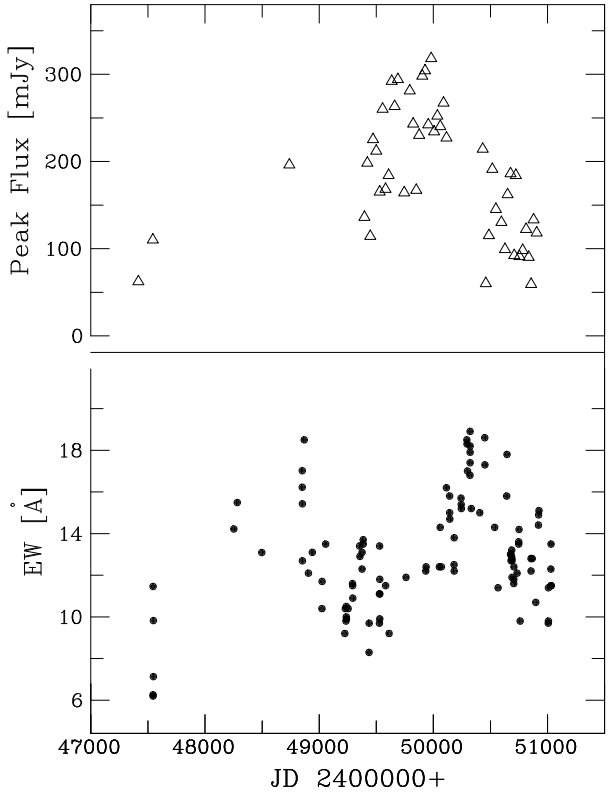}
\caption[]
     {Long term behavior of the H$\alpha$ equivalent width of \lsi. The
upper panel represents the radio outburst peak flux from Gregory (1999)}
\label{EWJD}
\end{figure}


The H$\alpha$ parameters folded on the radio period, are displayed in
Fig.~\ref{P26} and illustrate the general trend of the H$\alpha$ orbital
modulation. We used for this plot the 26.4917 d period value from radio
wavelengths as the best estimate of the orbital cycle currently available
(Gregory et al. 1999). In the panel of the radial velocities only the
Rozhen and Crimean data are included. All data from Table 1 as well as the
data from Paredes et al. (1994) are incorporated in other panels. The
ratio $FWHM(B)/FWHM(R)$ achieves maximal values at radio phases $0.3-0.6$.
The equivalent width ratio has one well defined maximum at phases
$0.25-0.4$. In the $B/R$ ratio at least one maximum is visible at phase
0.25, which coincides with the maximums of $FWHM(B)/FWHM(R)$ and
$EW(B)/EW(R)$. A second maximum is visible at phase 0.8. It deserves
noting that the appearance of this maximum is mainly due to 5 old points,
which are very noisy and the presence of this maximum must be confirmed by
future observations. These results proof the previous suggestions by
Paredes et al. (1994) and Zamanov et al. (1996) that an orbital
variability in the line ratios exists.

\begin{figure}[htb]
\mbox{}
\vspace{15.7cm}
\includegraphics{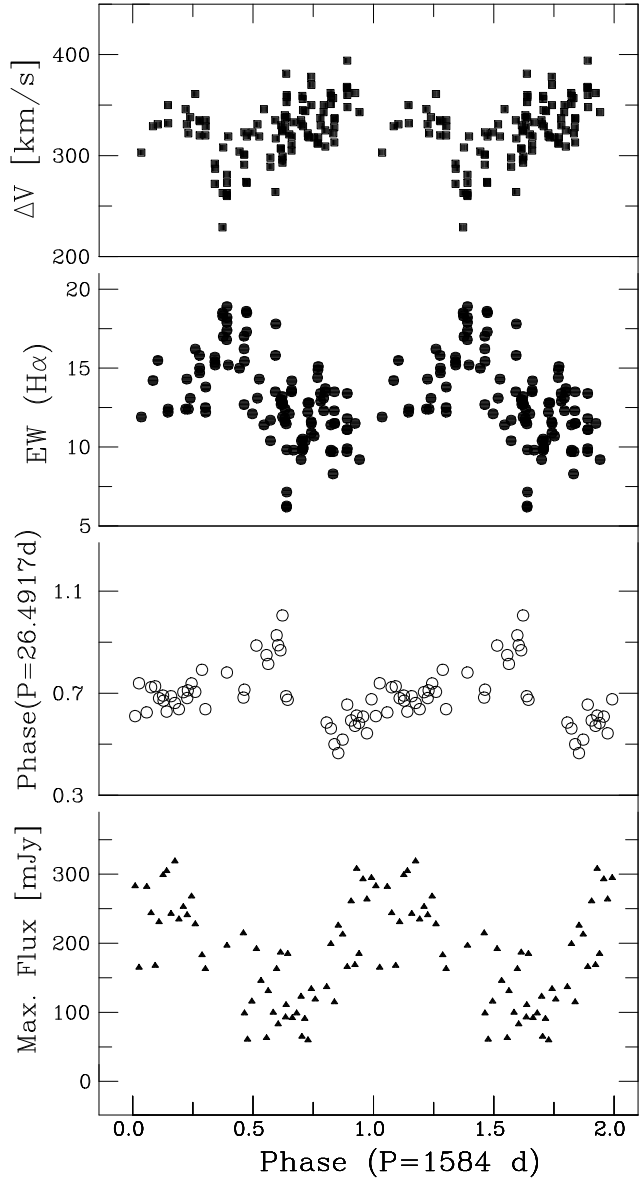}
\caption[]
     {H$\alpha$ and radio parameters of \lsi\ folded on the 1584 day
period with zero phase set at JD2443366.775. From up to down are plotted
the hump separation of the H$\alpha$, the total $EW($H$\alpha)$, the
phase of the radio maximum with respect to the 26.5 day period, and the
maximum radio flux}
\label{P1600}
\end{figure}


\subsection{Detection of a possible $\sim4$ yr modulation in H$\alpha$}

Fig.~\ref{EWJD} represents the long term evolution of the $EW($H$\alpha)$,
which varies over the range 8-20 \AA. These are real variations in the
H$\alpha$ emitted flux, because the photometric observations of Mendelson
\& Mazeh (1995) have shown that the $R$-band flux is stable with maximal
deviation of the points from the average $\pm 0.03$ magnitudes only. Two
maxima in the $EW($H$\alpha)$ are visible, at about JD 2448850 and JD
2450300. The time interval between these two maxima is $\sim 1500$ d,
which is remarkably close to the 1605 day period of the radio modulation
as estimated by Mart\'{\i} (1993) and 1584 days by Gregory et al. (1999).

We have performed the period search for long term modulation. Although the
values obtained are not well constrained they also point towards the four
year modulation.

In Fig.~\ref{P1600} the H$\alpha$ line parameters $EW($H$\alpha)$ and
$\Delta V = V_r(R)-V_r(B)$ are plotted together with the radio parameters
- the phase of the radio outburst and the maximum radio flux. The radio
data are taken from Gregory (1999). All data are folded on the 1584 day
period. Although the scatter of the points is considerable, the trend of
this plot strongly suggests that the $\sim4$ yr modulation in the
H$\alpha$ line is very likely to be real.

\subsection{Other period and correlation searches}

In addition to the $EW($H$\alpha)$, we also tried to find evidence for
long term periodicities (100-2000 days) in other line parameters. The
presence of $B/R$ and related variability cycles, with periods between 4
months and 3 years, is currently regarded as a common characteristic of
Be/X-ray binaries (Negueruela et al. 1998). Similar variability cycles
exist in isolated Be stars, with longer periods ranging from 2 years to
decades, with an average of 7 years. In those cases, the periodic
variability is interpreted as due to the propagation of global density
waves in the circumstellar disk (Okazaki 1997). Unfortunately, the results
of this search turned out to be negative. Thus, the suggestion of a
$\sim4$ yr H$\alpha$ modulation remains present in the $EW($H$\alpha)$ and
the hump separation measurements.

Possible correlations between the $EW($H$\alpha)$ and the radio outburst
amplitude were investigated as well. We used for this purpose the
NSF-NRAO-NASA Green Bank Interferometer data analyzed by Gregory (1999).
For every H$\alpha$ data point, the closest radio maximum was selected and
correlations were searched for. Again, none was found in reliable way. We
do not rule out that all these negative results are a consequence of the
considerable scatter of the points, not to say our still short time
baseline for long term period studies. It deserves noting that the radio
observations are missing just about the time of the $EW($H$\alpha)$
maximum at JD2450300 and the H$\alpha$ observations are very sparse at the
maximum of the radio flux modulation (see Fig.~\ref{EWJD}).

\section{Discussion}

\subsection{The 26.5 day period and the Be envelope}

Our spectroscopic results have provided a direct connection between the
periodic radio outbursts of \lsi\ and its Be disk environment. This
connection is based on the suggestive evidence that some of the H$\alpha$
emission line parameters also vary with the same 26.5 d and possibly
$\sim4$ yr periods previously found at radio wavelengths. Therefore, the
outburst models based on the interaction between the compact object and
the star Be envelope are clearly supported.

In this context, it appears likely that the 26.5 d H$\alpha$ variations
have to be interpreted as the neutron star perturbing the Be envelope as
it was discussed by Zamanov et al. (1996). However, other alternatives
cannot be completely ruled out. For instance, the detected 26.5 d period
in $B/R$ and other ratios could also be caused by the propagation of a
global density wave, as it happens in many Be X-ray binaries and isolated
Be stars. This scenario would then imply the shortest propagation period
of the global oscillation detected so far, and the first system in which
the propagation of the density wave is synchronized with the orbital
period of the neutron star.

\subsection{What causes the $\sim4$ yr modulation in \lsi?}

The H$\alpha$ observations presented here can be also used to discriminate
between the two previously proposed models for the $\sim4$ yr modulation
in the radio outburst amplitude. In the \lsi\ case, it appears that the
precessing jet model should be ruled out in favor of those based on
quasi-cyclic Be-star envelope variations. Indeed, the most likely
explanation of the $EW($H$\alpha)$ changes, that we observe in a $\sim4$
yr time scale, is a variability of the mass loss rate of the Be star.

To better assess this issue, it is instructive to obtain a estimation of
how much the mass loss rate should change to account for the observed
variations. We will use for this purpose the formulae of V\"ogel (1993)
for an optically thin emission line. Of course, the H$\alpha$ line is
probably not thin, so it must be considered as a rough estimation of the
mass loss rate variability. The flux of a wind line determined by
recombination can be represented as
 \begin{equation}
  F = \frac {h \nu} {4\pi d^2} \: \int_{\cal{V}}
  \alpha_{\rm eff}(n_{\rm e},T_{\rm e})\:n_{\rm e}\:n_{\rm ion}\:d\cal{V},
 \end{equation}
where $d$ is the distance to the star, $n_{\rm e}$ is the electron
density, $n_{\rm ion}$ the density of the ion considered, $\alpha_{\rm
eff}(n_{\rm e},T_{\rm e})$ is the effective recombination coefficient of
the line, and $\cal{V}$ the emitting volume. In the circumstellar disk of
\lsi\ the density follows the power law $\rho (r)= \rho _0
(r/R_*)^{-3.25}$ (Waters et al. 1988), so the \lsi\ flux in H$\alpha$ can
be expressed as
  \begin{equation}
   F=A\: \dot M_{\rm loss} ^2 \: \left[ 1-\left(\frac{R_{\rm in}}{R_{\rm out}}\right)^{3.5} \right]
  \end{equation}
where $\dot M_{\rm loss}$ is the mass loss rate of the Be star, $R_{\rm
in}$ and $R_{\rm out}$ are the inner and outer radius of the H$\alpha$
emitting disk, and the constant $A$ includes the average recombination
coefficient, the helium abundance and so on (for more details e.g.
V\"ogel, 1993; Tomov et al. 1998). We adopt that $R_{\rm in}=R_*$ and for
Keplerian disk $\Delta V/(2\,v\,\sin{i})=(R_{\rm out}/R_*)^{-1/2}$
(Hanuschik 1988), where $\Delta V$ and $v\,\sin{i}$ are the same as in
Section 3.1. Since $F\propto EW$, the observed H$\alpha$ parameters of
\lsi\ correspond to
  \begin{equation}
   \dot M_{\rm loss,max}/\dot M_{\rm loss,min} \simeq 1.5.
  \end{equation}
In this way, the variability in $EW($H$\alpha)$ can be explained by
suggesting that the mass loss rate varies by a plausible amount of $\pm
25$ per cent over its average value. This mechanism will also imply
variations of the density over the same range.

The interaction of a neutron star with the Be star wind is critical in all
suggested models for the radio outbursts of \lsi. The variability of the
mass loss rate of the Be star causes a variability of the mass accretion
rate onto the neutron star in case of accretion onto the magnetosphere. In
the model of young radio pulsar, the variability of the Be star mass loss
changes the position of the shock front. In ejector-propeller model it
affects the moments of switching of the Ejector - at lower mass loss rate
we expect the ejector to switch on earlier and the radio outburst to peak
at an earlier orbital phase than at higher mass loss rate.

In Fig.~\ref{P1600}, it appears that the H$\alpha$ measurements are not in
phase nor in anti phase with the radio data. This can be due to the bad
distribution of the observations as it was noted above or to a real phase
shift between the H$\alpha$ and radio parameters. At this moment we are
only able to say that the 1600 day modulation detected in the phase
variations of the 26.5 day periodic radio outbursts and the outburst peak
flux density is visible in the H$\alpha$ emission line too.

In any case, it is clear that additional theoretical work and
spectroscopic observations will be necessary to find out what is the exact
relationship between the radio outbursts and the circumstellar disk.

\subsection{Expectations at other wavelengths}

Finally, if the $\sim4$ yr modulation in radio and H$\alpha$ is due to
changes in the circumstellar disk density, it can be probably detected as
well in other wavelength domains. For instance, we can expect it to be
visible in the infrared bands, where the circumstellar disk contributes
with an excess above the stellar continuum. The X-rays and $\gamma$-rays
may also be modulated every $\sim4$ yr too, since the high energy emission
of the neutron star depends on the density of the surrounding matter with
which it interacts.

\section{Conclusions}

The main results from our H$\alpha$ spectroscopic observations of \lsi\
are:

\begin{enumerate}

\item The circumstellar disk in \lsi\ is among the densest of Be stars.

\item The 26.5 day period is visible in the H$\alpha$ emission line. This
is the first time that a clear stable periodicity is detected in the
H$\alpha$ emission line of a Be/X-ray binary in connection with the
orbital motion of the compact object.

\item Evidences of a $\sim4$ yr modulation in the H$\alpha$ of \lsi\ have
been found. This favours the $\sim4$ yr cycle in the radio outbursts as
being related to variations of the circumstellar disk. The model of
precessing jets should be consequently ruled out.

\item The $\sim4$ yr modulation is also likely to be present at the
infrared, X-ray and $\gamma$-ray domains.

\end{enumerate}

\begin{acknowledgements}

RZ acknowledges support from the Bulgarian NSF (MUF-05/96).
JMP and JM acknowledge partial support by DGICYT (PB97-0903).
MR is supported by a fellowship from CIRIT (Generalitat de Catalunya, ref.
1999 FI 00199). JM is also partially supported by Junta de Andaluc\'{\i}a
(Spain).

\end{acknowledgements}

\begin{table*}
\caption[]{H$\alpha$ line parameters of \lsi}

\begin{tabular}{c@{\hspace{1ex}}l@{\hspace{1ex}}cccccccccc}
\hline
 Date & JD$-$2400000 & $-EW($H$\alpha)$ & $V_r(R)$  & $V_r(B)$ & $V_r$(dip) & $EW(B)/EW(R)$ & $B/R$ & $FWHM(B)$ & $FWHM(R)$ \\
(yymmdd) &           &     (\AA)      &(km s$^{-1}$)&(km s$^{-1}$)&(km s$^{-1}$)&         &    &(\AA) &(\AA)  \\ [1ex]
\hline
          &               &     &     &     &     &     &     &      \\
Ro920903  & 48869.50& 18.5&  110& -209&  -69& 0.68& 0.77& 4.55& 6.68 \\
Ro921009  & 48905.48& 12.1&  130& -193&  -46& 0.61& 0.91& 3.65& 5.27 \\
Ro930205  & 49024.27& 10.4&   83& -215&  -88& 0.60& 0.73& 5.55& 5.52 \\
Ro930206  & 49025.30& 11.7&  105& -184&  -76& 0.68& 0.68& 5.57& 5.15& \\
Ro930825  & 49225.45&  9.2&  132& -203&  -61& 0.93& 0.96& 5.43& 5.75& \\
Ro930829  & 49228.50& 10.4&  139& -192&  -26& 1.35& 1.13& 7.47& 5.76& \\
Ro930905  & 49235.49&  9.8&  101& -258& -109& 0.88& 0.70& 5.85& 6.27& \\
Ro930907  & 49237.50& 10.5&  127& -229&  -84& 0.96& 0.78& 6.02& 5.86& \\
Ro930908  & 49238.53& 10.0&  117& -228&  -94& 0.82& 0.81& 5.61& 6.17& \\
Ro930909  & 49239.50&  9.9&  109& -220&  -96& 0.92& 0.83& 5.82& 6.09& \\
Ro931031  & 49292.30& 11.6&  139& -239&  -78& 0.72& 0.88& 3.42& 4.68& \\
Ro931101  & 49293.32& 11.5&  128& -242&  -80& 0.78& 0.91& 3.57& 4.35& \\
Ro931102  & 49294.49& 10.9&  116& -234&  -70& 0.86& 0.89& 3.61& 4.12& \\
Ro940101  & 49354.22& 13.4&  129& -208&  -53& 0.85& 0.93& 3.00& 3.96& \\
Ro940103  & 49356.28& 12.9&  127& -203&  -48& 0.84& 0.87& 3.40& 3.74& \\
Ro940122  & 49375.50& 13.1&  123& -220&  -63& 0.88& 0.92& 3.37& 3.79& \\
Ro940123  & 49376.27& 12.3&  128& -222&  -63& 0.87& 0.88& 3.13& 3.49& \\
Ro940201  & 49385.34& 13.7&   85& -224&  -72& 0.87& 0.94& 3.40& 3.68& \\
Ro940202  & 49386.27& 13.5&  123& -202&  -41& 0.98& 0.97& 3.65& 3.36& \\
Ro940624  & 49528.49& 11.1&  120& -247&  -80& 0.81& 0.69& 4.03& 3.10& \\
Ro940625  & 49529.49& 13.4&  137& -257& -100& 0.65& 0.61& 3.59& 4.30& \\
Ro940626  & 49530.49& 11.8&  132& -235&  -75& 0.76& 0.78& 3.27& 4.44& \\
Ro940627  & 49531.47& 11.1&  111& -237&  -91& 0.82& 0.93& 3.08& 4.54& \\
Ro940818  & 49582.54& 11.5&  140& -222&  -54& 0.87& 0.87& 3.04& 3.98& \\
Ro960407  & 50181.30& 12.5&   91& -229&  -63& 0.98& 1.10& 3.17& 3.55& \\
Ro960408  & 50182.30& 12.2&  104& -231&  -49& 1.10& 1.03& 3.73& 3.35& \\
Ro960409  & 50183.36& 13.8&  102& -227&  -46& 1.20& 1.01& 4.35& 3.54& \\
Ro960608  & 50242.52& 15.7&   68& -204&  -91& 0.88& 0.78& 3.49& 3.09& \\
Ro960609  & 50243.52& 15.4&   82& -210&  -75& 0.90& 0.74& 3.29& 2.88& \\
Ro960610  & 50244.50& 15.2&   73& -214&  -90& 0.78& 0.71& 3.46& 2.86& \\
Ro960729  & 50293.55& 18.5&   88& -141&  -39& 1.08& 0.95& --- & --- & \\
Ro960730  & 50294.56& 18.3&  110& -153&  -42& 0.88& 0.88& 4.08& 3.28& \\
Ro960803  & 50298.59& 17.0&  111& -197&  -55& 0.91& 0.82& 3.57& 3.17& \\
Ro960824  & 50320.58& 16.8&   70& -191&  -82& 0.99& 0.82& 4.27& 3.22& \\
Ro960825  & 50321.53& 18.2&   65& -195&  -88& 0.89& 0.76& 3.89& 3.71& \\
Ro960826  & 50322.36& 18.9&   68& -195&  -88& 0.93& 0.74& 4.29& 3.14& \\
Ro960826  & 50322.49& 17.4&   78& -195&  -81& 0.88& 0.69& 4.19& 3.11& \\
Ro960827  & 50322.57& 17.9&   81& -200&  -80& 0.92& 0.68& 4.00& 3.11& \\
Ro970101  & 50450.36& 18.6&   94& -180&  -68& 0.85& 0.78& 4.65& 3.40& \\
Ro970102  & 50451.33& 17.3&   74& -199&  -74& 0.91& 0.79& 4.00& 3.90& \\
Ro970328  & 50536.26& 14.3&   85& -234&  -93& 0.90& 0.68& 4.13& 3.08& \\
Ro970426  & 50565.31& 11.4&  102& -244&  -86& 0.93& 0.99& 3.21& 4.05& \\
Ro970711  & 50641.47& 15.8&   84& -180&  -40& 0.90& 0.61& 3.93& 3.21& \\
Ro970713  & 50643.52& 17.8&  102& -215&  -75& 0.93& 0.72& 4.53& 2.99& \\
Ro970815  & 50676.50& 13.0&   96& -211&  -74& 0.71& 0.87& 3.10& 3.62& \\
Ro970816  & 50677.36& 13.0&   88& -219&  -87& 0.81& 0.86& 3.23& 3.32& \\
Ro970819  & 50680.56& 12.7&  106& -224&  -73& 0.83& 0.88& 2.85& 3.23& \\
Ro970824  & 50685.49& 12.9&  107& -217&  -48& 1.08& 1.00& 3.47& 3.28& \\
Ro970825  & 50686.47& 13.2&   85& -211&  -45& 1.18& 0.98& 4.49& 2.88& \\
Ro970826  & 50687.55& 12.8&   79& -214&  -57& 1.09& 0.96& 3.73& 3.78& \\
Ro970827  & 50688.59& 11.9&   97& -228&  -70& 0.88& 0.76& 3.71& 3.49&  \\
Ro970828  & 50689.60& 12.7&  102& -197&  -48& 0.93& 0.76& 4.02& 3.34&  \\
Ro970911  & 50703.38& 11.6&  110& -227&  -77& 0.78& 0.83& 3.07& 3.54& \\
Ro970912  & 50704.56& 11.8&  123& -217&  -76& 0.65& 0.76& 2.77& 3.80& \\
Ro970913  & 50705.54& 12.4&  122& -207&  -66& 0.71& 0.70& 3.04& 3.41& \\
          &               &     &     &     &     &     &     &      \\
\hline
\end{tabular}
\end{table*}
\newpage

\addtocounter{table}{-1}

\begin{table*}
\caption[]{Continuation}

\begin{tabular}{c@{\hspace{1ex}}l@{\hspace{1ex}}cccccccccc}
\hline
 Date & JD$-$2400000 & $-EW($H$\alpha)$ & $V_r(R)$  & $V_r(B)$ & $V_r$(dip) & $EW(B)/EW(R)$ & $B/R$ & $FWHM(B)$ & $FWHM(R)$ \\
(yymmdd) &           &     (\AA)      &(km s$^{-1}$)&(km s$^{-1}$)&(km s$^{-1}$)&         &    &(\AA) &(\AA)  \\ [1ex]
\hline
          &               &     &     &     &     &     &     &      \\
Ro971010  & 50732.40& 12.1&  124& -197&  -59& 0.74& 0.87& 3.25& 3.60& \\
Ro980209  & 50854.21& 12.2&   97& -225&  -66& 0.94& 0.91& 3.12& 3.32& \\
Ro980210  & 50855.22& 12.8&   99& -220&  -77& 0.83& 0.90& 3.29& 3.63& \\
Ro980219  & 50864.23& 12.8&  106& -212&  -80& 0.64& 0.70& 3.14& 3.70& \\
Ro980323  & 50896.28& 10.7&  100& -218&  -58& 1.20& --- & 3.42& 3.62& \\
Ro980415  & 50919.27& 14.4&  104& -208&  -71& 0.85& 0.84& 2.89& 3.58& \\
Ro980417  & 50921.59& 14.9&   97& -215&  ---& 1.03& --- & 3.17& 3.68& \\
Ro980419  & 50923.59& 15.1&   93& -225&  -78& 0.89& 0.86& 3.45& 3.46& \\
Ro980710  & 51005.56&  9.7&  122& -240&  -48& 1.05& 0.89& 4.88& 3.53& \\
Ro980711  & 51006.51& 11.4&  124& -227&  -49& 1.25& 1.00& 5.36& 3.54& \\
Ro980712  & 51007.49&  9.8&  128& -231&  -43& 1.06& 1.04& 4.41& 3.55& \\
Ro980802  & 51028.56& 12.3&  122& -206&  -52& 0.84& 0.92& 3.06& 3.16& \\
Ro980803  & 51029.53& 11.5&  124& -212&  -48& 0.92& 0.92& 3.37& 3.22& \\
Ro980804  & 51030.52& 13.5&  117& -196&  -39& 1.08& 0.96& 4.28& 3.39& \\
Ro980805  & 51031.56& 11.5&  114& -223&  -58& 0.85& 0.80& 3.68& 3.44& \\
          &               &     &     &     &     &     &     &       \\
SV921113  & 48939.80& 13.1&  111& -220&  -62& 0.81& 0.77& 3.62& 3.29& \\
SV930310  & 49056.70& 13.5&   93& -242&  -95& 0.85& 0.96& 3.30& 3.98& \\
SV930923  & 49253.90& 10.4&   96& -248&  -65& 1.06& 1.14& 3.94& 4.01& \\
SV940326  & 49437.34&  8.3&   90& -267& -101& 0.72& 1.00& 2.72& 3.96& \\
SV940327  & 49438.37&  9.7&   66& -256&  ---& 0.87& 0.94& 3.31& 4.46& \\
SV940624  & 49528.69&  9.7&  147& -221&  -44& 0.78& 0.66& 3.61& 3.09& \\
SV940626  & 49530.69&  9.9&  140& -220&  -70& 0.76& 0.81& 2.76& 4.45& \\
SV940916  & 49611.61&  9.2&  130& -213&  -54& 0.93& 1.07& 3.24& 3.92& \\
SV950212  & 49760.43& 11.9&  168& -135&   23& 1.00& 0.95& 3.81& 3.69& \\
SV950804  & 49933.73& 12.2&  126& -224&  -53& 0.91& 0.88& 3.12& 2.81& \\
SV950806  & 49935.70& 12.4&  128& -204&  -62& 0.74& 0.86& 2.98& 3.10& \\
SV960228  & 50141.43& 15.8&  132& -203&  -56& 0.70& 0.80& 2.87& 3.86& \\
SV960229  & 50142.39& 15.0&  131& -203&  -55& 0.70& 0.88& 2.88& 3.92& \\
SV960301  & 50143.40& 14.7&  122& -198&  -53& 0.71& 0.96& 2.67& 4.20& \\
SV971026  & 50747.52& 13.6&   91& -214&  -73& 0.84& 0.70& 3.12& 3.01& \\
SV971026  & 50747.66& 13.5&  103& -208&  -73& 0.85& 0.75& 3.19& 3.15& \\
SV971028  & 50749.48& 14.2&  113& -206&  -54& 0.86& 0.89& 3.08& 3.28& \\
          &               &     &     &     &     &     &     &       \\
Cr951129  & 50051.37& 12.4&  125& -221&  -85& 0.64& 0.63& 3.61& 3.67& \\
Cr951207  & 50059.21& 14.3&  124& -207&  -66& 0.73& 0.71& 3.00& 3.68& \\
Cr951217  & 50069.22& 12.4&  134& -188&  -49& 0.79& 0.97& 2.60& 3.13& \\
Cr960131  & 50114.32& 16.2&  160& -201&  -85& 0.50& 0.78& 2.51& 6.40& \\
Cr960904  & 50331.47& 15.2&  120& -199&  -65& 0.65& 0.78& 3.31& 3.52& \\
Cr961118  & 50406.33& 15.0&  109& -195&  -65& 0.71& 0.76& 2.75& 3.65& \\
Cr971107  & 50760.41&  9.8&  124& -209&  -45& 0.91& 1.01& 3.12& 3.62& \\
          &               &     &     &     &     &     &     &       \\
\hline
\end{tabular}
~\\
Ro: ``Rozhen'' Observatory; SV: Southampton \& Valencia Universities
collaboration; Cr: Crimean Astrophysical Observatory.

\end{table*}

\end{document}